\documentclass{webofc}
\usepackage[utf8]{inputenc}
\usepackage[labelfont=bf]{caption}
\usepackage{subcaption}
\usepackage{listings}
\usepackage{hyperref}
\usepackage{graphicx}
\usepackage[draft]{minted}
\usepackage{graphicx}
\usepackage{tikz}
\usepackage{multicol}

\usetikzlibrary{calc}

\usepackage[varg]{txfonts}

\newminted{cpp}{gobble=2, numbersep=3pt, escapeinside=||, xleftmargin=5mm, framesep=2mm, frame=leftline, baselinestretch=1, fontsize=\scriptsize, linenos=true}
\newmintinline{cpp}{}

\makeatletter
\AtBeginEnvironment{minted}{\dontdofcolorbox}
\def\dontdofcolorbox{\renewcommand\fcolorbox[4][]{##4}}
\makeatother

\usetikzlibrary{shadows, arrows, fit, positioning, matrix, shapes}
\usetikzlibrary{decorations.markings}

\definecolor{my_yellow}{RGB}{255, 253, 217}
\definecolor{my_orange}{RGB}{255, 127, 0}
\definecolor{my_lightblue}{RGB}{105, 186, 249}
\definecolor{my_purple}{RGB}{150, 154, 219}
\definecolor{my_green}{RGB}{90, 194, 160}

\tikzset {
  bigbox/.style = {draw, thick, fill=gray!10, rounded corners, rectangle},
  box/.style = {draw, thick, minimum height=0.8cm, minimum width=1.5cm, rounded corners, rectangle, fill=white, anchor=south},
  model/.style = {draw, thick, fill=white, text centered, minimum height=3em, minimum width=4em, rounded corners, drop shadow},
  user/.style = {draw, thick, ellipse, fill=white, text centered, minimum height=3em, minimum width=5em, drop shadow},
  line/.style = {->, thick, color=black, shorten <=2pt, shorten >=2pt, >=stealth'},
  dashedline/.style = {->, thick, dashed, color=black, shorten <=2pt, shorten >=2pt, >=stealth'},
  plain/.style = {minimum width=1em},
  arcnode/.style 2 args={
    decoration={
                 raise=#1,             
                 markings,   
                 mark=at position 0.5 with {\node[inner sep=0] {#2};}
            },
            postaction={decorate}
    }
}

\usetikzlibrary{shadows, calc}

\def\shadowshift{3pt,-3pt}
\def\shadowradius{6pt}

\colorlet{innercolor}{black!30}
\colorlet{outercolor}{gray!05}

\newcommand\drawshadow[1]{
  \begin{pgfonlayer}{shadow}
    \shade[outercolor,inner color=innercolor,outer color=outercolor] ($(#1.south west)+(\shadowshift)+(\shadowradius/2,\shadowradius/2)$) circle (\shadowradius);
    \shade[outercolor,inner color=innercolor,outer color=outercolor] ($(#1.north west)+(\shadowshift)+(\shadowradius/2,-\shadowradius/2)$) circle (\shadowradius);
    \shade[outercolor,inner color=innercolor,outer color=outercolor] ($(#1.south east)+(\shadowshift)+(-\shadowradius/2,\shadowradius/2)$) circle (\shadowradius);
    \shade[outercolor,inner color=innercolor,outer color=outercolor] ($(#1.north east)+(\shadowshift)+(-\shadowradius/2,-\shadowradius/2)$) circle (\shadowradius);
    \shade[top color=innercolor,bottom color=outercolor] ($(#1.south west)+(\shadowshift)+(\shadowradius/2,-\shadowradius/2)$) rectangle ($(#1.south east)+(\shadowshift)+(-\shadowradius/2,\shadowradius/2)$);
    \shade[left color=innercolor,right color=outercolor] ($(#1.south east)+(\shadowshift)+(-\shadowradius/2,\shadowradius/2)$) rectangle ($(#1.north east)+(\shadowshift)+(\shadowradius/2,-\shadowradius/2)$);
    \shade[bottom color=innercolor,top color=outercolor] ($(#1.north west)+(\shadowshift)+(\shadowradius/2,-\shadowradius/2)$) rectangle ($(#1.north east)+(\shadowshift)+(-\shadowradius/2,\shadowradius/2)$);
    \shade[outercolor,right color=innercolor,left color=outercolor] ($(#1.south west)+(\shadowshift)+(-\shadowradius/2,\shadowradius/2)$) rectangle ($(#1.north west)+(\shadowshift)+(\shadowradius/2,-\shadowradius/2)$);
    \filldraw ($(#1.south west)+(\shadowshift)+(\shadowradius/2,\shadowradius/2)$) rectangle ($(#1.north east)+(\shadowshift)-(\shadowradius/2,\shadowradius/2)$);
  \end{pgfonlayer}
}

\newsavebox\mybox
\newlength\mylen

\newcommand\shadowimage[2][]{%
\setbox0=\hbox{\includegraphics[#1]{#2}}
\setlength\mylen{\wd0}
\ifnum\mylen<\ht0
\setlength\mylen{\ht0}
\fi
\divide \mylen by 120
\def\shadowshift{\mylen,-\mylen}
\def\shadowradius{\the\dimexpr\mylen+\mylen+\mylen\relax}
\begin{tikzpicture}
  \node[fill=white, rectangle, rounded corners, anchor=south west, inner sep=0] (image) at (0,0) {\includegraphics[#1]{#2}};
  \drawshadow{image}
\end{tikzpicture}}

\pgfdeclarelayer{background}
\pgfdeclarelayer{foreground}
\pgfdeclarelayer{shadow}
\pgfsetlayers{shadow,background,main,foreground}

\newcommand*\circled[1]{\tikz[baseline=(char.base)]{
            \node[shape=circle,draw,inner sep=2pt] (char) {#1};}}

\title{Automatic Differentiation in ROOT}
        \author{
            \firstname{Vassil}
            \lastname{Vassilev}\inst{1}\fnsep\thanks
            {\email{vvasilev@cern.ch}}
            \firstname{Aleksandr}
            \lastname{Efremov}\inst{1}\fnsep\thanks
            {\email{xefremale@gmail.com}}
            \firstname{Oksana}
            \lastname{Shadura}\inst{2}\fnsep\thanks
            {\email{oksana.shadura@cern.ch}}
}

\institute{Princeton University, Princeton, New Jersey 08544, United States
\and University of Nebraska Lincoln, 1400 R St, Lincoln, NE 68588, United States
}

\begin{document}

\abstract{%
In mathematics and computer algebra, automatic differentiation (AD) is a set of techniques to evaluate the derivative of a function specified by a computer program. AD exploits the fact that every computer program, no matter how complicated, executes a sequence of elementary arithmetic operations (addition, subtraction, multiplication, division, etc.), elementary functions (exp, log, sin, cos, etc.) and control flow statements. AD takes source code of a function as input and produces source code of the derived function. By applying the chain rule repeatedly to these operations, derivatives of arbitrary order can be computed automatically, accurately to working precision, and using at most a small constant factor more arithmetic operations than the original program.

This paper presents AD techniques available in ROOT, supported by Cling, to produce derivatives of arbitrary C/C++ functions through implementing source code transformation and employing the chain rule of differential calculus in both forward mode and reverse mode. We explain its current integration for gradient computation in TFormula. We demonstrate the correctness and performance improvements in ROOT's fitting algorithms.
}
\maketitle
\section{Introduction} \label{intro}

Accurate and efficient computation of derivatives is vital for a wide variety of computing applications, including numerical optimization, solution of nonlinear equations, sensitivity analysis, and nonlinear inverse problems. Virtually every process could be described with a mathematical function, which  can be thought of as an association between elements from different sets. Derivatives track how a varying quantity depends on another quantity, for example how the position of a planet varies as time varies.

Derivatives and gradients (vectors of partial derivatives of multivariable functions) allow us to explore the properties of a function and thus the described process as a whole. Gradients are an essential component in gradient-based optimization methods, which have become more and more important in recent years, in particular with its application training of (deep) neural networks \cite{baydin2017automatic}.

Several different techniques are commonly used to compute the derivatives of a given function, either exactly or approximately \cite{gebremedhin2020introduction, baydin2017automatic, griewank2008evaluating}. The most prevalent techniques are:

\begin{itemize}
    \item \emph{Numerical differentiation}, based on the \emph{finite difference} method, provides a way to evaluate derivatives approximately. While simple, numerical differentiation can be  slow (the run-time complexity grows linearly with the number of input variables) and may have problems with accuracy due to round-off and truncation errors.
    \item \emph{Symbolic differentiation}, based on transformations of symbolic expressions of functions, provides exact closed-form expressions for the derivatives. It faces difficulties when the function to be differentiated is not available in a closed form, which is often the case for computer programs which may contain control flow. Symbolic differentiation can produce derivative expressions that are computationally expensive to evaluate due to difficulties in exploiting common subexpressions.
    \item \emph{Automatic differentiation} (AD)  computes derivatives accurately to the precision of the original function, supports control flow and uses at most a small constant factor more time and space than it takes to evaluate the original function, at the expense of increased implementation complexity and introducing more software dependencies.
\end{itemize}
Numerical and symbolic differentiation methods are slow at computing gradients of functions with many input variables, as is often needed for gradient-based optimization algorithms. Both methods have problems calculating higher-order derivatives, where the complexity and errors due to numerical precision increase. Automatic differentiation largely avoids the problems of numerical and symbolic differentiation.

In this paper, we describe the implementation of automatic differentiation techniques in ROOT, which is the data analysis framework broadly used High-Energy Physics~\cite{Brun1997ROOT}. This implementation is based on Clad~\cite{vassilev2015clad, cladgithub}, which is an automatic differentiation plugin for computation expressed in C/C++.




\section{Background} \label{background}
Here, we briefly discuss main algorithmic and implementation principles behind AD. An in-depth overview and more formal description can be found in ~\cite{gebremedhin2020introduction} and~\cite{griewank2008evaluating}, respectively.
\subsection{AD and its Modes}
AD is based on the decomposition of the procedure (e.g. a source code that computes the original function) into a sequence of simple mathematical operations (e.g. $+, -, *, /, \sin, \cos, \exp$) that can be expressed using a series of intermediate results. Subsequently, derivatives of every intermediate result are evaluated and combined via the chain rule of calculus to obtain the derivatives of the whole sequence. The control flow (e.g. branches, loops) can be incorporated by differentiating the control flow of the original function during the derivative evaluation. Two main modes of AD, which differ in the order of application of the chain rule, are used:
\begin{itemize}
    \item \emph{Forward mode} operates in a top-down approach and computes the derivative of every intermediate result with respect to a single selected input variable of the function. As soon as a final result of the function is reached, the partial derivative with respect to the selected input is available.
    A single evaluation of the forward mode can only compute partial derivatives with respect to a single input variable. Thus, when the whole gradient is required, forward mode must be invoked once per every input variable, leading to $m \cdot c_{F} \cdot n$ runtime complexity, where $m$ is the number of input variables, $n$ is the algorithmic complexity of the original function and $c_{F} < 3 $ is a small constant factor overhead of a single invocation of the forward mode~\cite{griewank2008evaluating}.
    \item \emph{Reverse mode} operates in a bottom-up approach and computes the derivative of a function's output with respect to every intermediate result. Once every input variable of the function is reached, the whole gradient of an output is available. Note that, independently on the number of input variables $N$, a single evaluation of the reverse mode is sufficient to get the whole gradient of a function's output, leading to $c_{R} \cdot n$ runtime complexity, where $n$ is the complexity of the original function and $c_{R} \leq 4$ is a small constant factor overhead \cite{griewank2008evaluating}. This is a huge advantage in settings with a single scalar output and many inputs, which is often the case in machine-learning problems where $N >> 10^6$ that makes the forward mode infeasible. As a disadvantage, reverse mode implementations are more complicated, and dynamic memory allocations may be required when dynamic control flow is involved. Depending on the original function, this may cause a single evaluation of the reverse mode to be somewhat slower compared to a single evaluation of the forward mode.
\end{itemize}
\subsection{AD Implementations}
AD techniques have been implemented in a variety of programming languages and paradigms, ranging from classical tools for Fortran~\cite{bischof1996adifor} and~C~\cite{bischof1997adic}, to recent active work on tools specific to machine-learning applications~\cite{paszke2017automatic, jax2018github}, and modern general-purpose programming languages~\cite{innes2019differentiable, swiftfortensorflow}.
We refer the reader to \texttt{www.autodiff.org} for a comprehensive list of available AD implementations for various languages.

In particular, several implementations exist for C++, e.g. \cite{autodiffcpp, hogan2014fast, walther2009getting}. Majority of implementations of AD fall into one of the two categories of implementation techniques:
\begin{itemize}
    \item Tools based on \emph{operator overloading} utilize features of programming languages like C++ and Python to define custom types and overload mathematical operators (e.g.~+,~-,~*,~/) and functions (e.g. $\exp, \sin, \cos$) on them. Such implementations are often based on custom AD-enabled types that wrap values of both the original and derivative functions and redefine operators to simultaneously act on original and derivative values. In C++, such tools are often implemented as a library that introduces templated differentiable types and corresponding mathematical operations. Then, functions called on the custom type return both original and derivative values.
    This is a powerful technique but has two primary limitations: legacy code and performance. Functions must be either polymorphic (templated) or explicitly defined on AD-enabled type to be differentiated. Differentiation of pre-existing source code using builtin types such as \texttt{double} and \texttt{float} is not possible. Users are required to use additional level of abstraction in the form of library-specific types instead of first-class language features. Moreover, the performance of the derivative generation can be suboptimal due to the C++ metaprogramming system which usually constructs deep template instantiation chains. Performance can be even more problematic when creating a higher order derivatives.
    \item Tools based on \emph{source transformation} analyze the source code of the original function and build another source code for the derivative function. Such techniques typically accept and generate any code using built-in features of the original language and do not require custom libraries. On the other hand, they require an additional pass over the source file to analyze and generate derivative code. Source transformation can fully utilize source-level optimizations and has reasonably good performance. Implementation is more complicated and it is problematic to achieve full coverage of C++ language features. While full integration with a compiler can make AD a first-class language feature that is transparent for the user, most current implementations for C++ are based on custom parsers that do not have full coverage of the vast variety of C++ language constructs and require a separate step before compilation.
\end{itemize}

\section{Architecture and Implementation} \label{architecture}
Automatic differentiation in ROOT is based on Clad \cite{vassilev2015clad, cladgithub}.
Clad is a source transformation AD tool for C++. It is based on LLVM compiler infrastructure \cite{LLVM:CGO04} and is implemented as a plugin for C++ compiler \emph{Clang}, which allows Clad to be transparently integrated into the compilation phase and to utilize large parts of the compiler. Clad relies on Clang's parsing and code generation functionality and can differentiate complicated C++ constructs. Clad supports both forward and reverse mode. It is available as a standalone Clang plugin that, when attached to the compiler, produces derivatives in the compilation phase.

On top of that, Clad is integrated directly into ROOT to provide AD functionality as an integral part of the framework. ROOT has a C++ interpreter \emph{Cling} \cite{Vasilev2012Cling}  which is built on the top of LLVM and Clang. This allows Clad to be attached to Cling as a plugin in a similar way as it can be attached to Clang. In this section, we discuss 1) architecture of Clad and its interaction with Cling; and 2) details of its integration into ROOT.

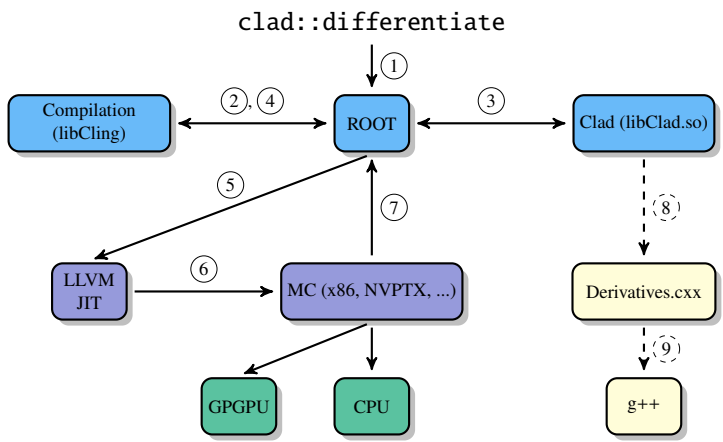
\begin{figure}[h!]
   \centering
   \begin{tikzpicture}[outer sep=0.05cm, node distance=0.8cm, scale=0.7, transform shape]
         
     \node[name=code] (code) {\Large{\texttt{clad::differentiate}}};
     \node[model, fill=my_lightblue, name=cling, below=1 cm of code] (cling) {ROOT};
     \node[model, fill=my_lightblue, text width=8em, name=clang, left=3cm of cling] (clang) {Compilation \\ (libCling)};
     \node[model, fill=my_lightblue, name=clad, right=3cm of cling] (clad) {Clad (libClad.so) };
     \node[model, fill=my_purple, text width=3em, name=llvm, below=2cm of clang] (llvm) {LLVM JIT};
     
     \node[model, fill=my_yellow, text width=7em, name=text_derivatives, below=2cm of clad] (text_derivatives) {Derivatives.cxx};
     \node[model, fill=my_yellow, text width=3em, name=gcc, below=1cm of text_derivatives] (gcc) {g++};
     
     \node[model, fill=my_purple, name=mc, below=2cm of cling] (mc) {MC (x86, NVPTX, ...)};
     \node[model, fill=my_green, name=cpu, below=1cm of mc] (cpu) {CPU};
     \node[model, fill=my_green, name=gpu, left=1cm of cpu] (gpu) {GPGPU};
 
     \draw[line, ->] (code.south) -- node[midway, right] {\circled{1}} (cling.north);
     \draw[line,<->] (cling.west) -- node[midway, above] {\circled{2}, \circled{4}} (clang.east);
     \draw[line,<->] (cling.east) -- node[midway, above] {\circled{3}} (clad.west);
     \draw[line,->] (cling.south) -- node[midway, above] {\circled{5}} (llvm.north);
     \draw[line,->] (llvm.east) -- node[midway, above] {\circled{6}} (mc.west);
     \draw[line,->] (mc.north) -- node[midway, right] {\circled{7}} (cling.south);
     \draw[line,->] (mc.south) -- (cpu.north);
     \draw[line,->] (mc.south) -- (gpu.north);

     \draw[dashedline,->] (clad.south) -- node[midway, right] {\circled{8}} (text_derivatives.north);
     \draw[dashedline,->] (text_derivatives.south) --  node[midway, right] {\circled{9}} (gcc.north);
   \end{tikzpicture}
   \caption{Information flow of Clad in ROOT}
   \label{fig:clad-architecture}
 \end{figure}


Clad operates on \emph{Clang AST} (abstract  syntax  tree) by analyzing the AST of the original function and generating the AST of the derivative. Clad provides two API functions: \cppinline{clad::differentiate} for forward mode and \cppinline{clad::gradient} for reverse mode, which can be used directly in the source code to mark a function for differentiation (see \cite{cladgithub} for more details on usage and code examples). 

The information flow of interactions with Cling during differentiation (Figure~\ref{fig:clad-architecture}) is:
\begin{itemize}
    \item A function is marked for differentiation with the C++ construct \texttt{clad::differentiate} or \texttt{clad::gradient} (step 1).
    \item Cling in ROOT performs incremental compilation and receives an abstract syntax tree (AST) representation of the code (step 2).
    \item Cling detects the differentiation marker and sends the AST of the original function to Clad, which transforms the AST to produce the AST of the derivative (step 3).
    \item Clad returns the derivative AST to Cling for code generation and execution by the low level LLVM primitives (steps 4, 5, 6, 7). \textit{Alternatively, if Clad was configured for non-interactive use, the generated AST can be converted to a C++ source code and written to a text file. The generated code then can be compiled with any C++ compiler (steps 8, 9).}
\end{itemize}

Inside of ROOT, interface functions \texttt{clad::differentiate} and \texttt{clad::gradient} are accessible via   \cppinline{#include <Math/CladDerivator.h>}. 
Clad is also directly integrated into the \cppinline{TFormula} class that encapsulates the concept of multidimensional mathematical functions in ROOT. \cppinline{TFormula} is a primitive in ROOT's math package which is connected to the Cling interpreter. In the context of \cppinline{TFormula}, Clad can differentiate functions available in the interpreter. The \cppinline{TFormula::GenerateGradientPar} method uses Clad to differentiate the underlying code of the formula with respect to its parameters and generate the code for the gradient. \cppinline{TFormula::GradientPar} method then evaluates the gradient at a specified point.

\section{Results} \label{experiments}
In this section, we empirically compare automatic differentiation (AD, our implementation based on Clad) and numerical differentiation (ND, based on finite difference method) in ROOT. We show that AD can drastically improve accuracy and performance of derivative evaluation, compared to ND. 
\subsection{Accuracy}
As stated in Section \ref{intro}, numerical differentiation may give imprecise results while AD computes the derivatives exactly. We show an example of a function where this difference is apparent: AD provides exact result while ND suffers from the loss of accuracy.

\vspace*{-0.25cm}
\begin{multicols}{2}
\noindent
\begin{equation}
p(x) = \frac{1}{\pi} \frac{\frac{1}{2} \Gamma}{x^2 + (\frac{1}{2} \Gamma)^2}\label{eq:1}
\end{equation}
\begin{equation}
    \frac{\partial{p(x)}}{\partial{\Gamma}}=-\frac{2}{\pi}\frac{\Gamma^2-4x^2}{(\Gamma^2+4x^2)^2}\label{eq:2}
\end{equation}
\end{multicols}
\vspace*{-0.25cm}

The function is the PDF of \emph{Breit-Wigner} distribution~(Eq.~\ref{eq:1}), whose derivative with respect to $\Gamma$ (Eq.~\ref{eq:2}) has critical points at $\Gamma=\pm{2x}$. In ROOT, the function is implemented as in~(Listing~\ref{lst:breitwigner_pdf}).

\vspace*{-0.15cm}
\begingroup
\begin{cppcode*}{linenos=false}
  inline double breitwigner_pdf(double x, double gamma, double x0 = 0) {
  double gammahalf = gamma/2.0;
  return gammahalf/(M_PI * ((x-x0)*(x-x0) + gammahalf*gammahalf));
}
\end{cppcode*}
\vspace*{-0.55cm}
\captionof{listing}{Breit-Wigner PDF implementation in ROOT}\label{lst:breitwigner_pdf}
\endgroup
\vspace*{-0.15cm}

When evaluating the derivative of \cppinline{breitwigner_pdf} with respect to \cppinline{gamma} at \cppinline{x=1}, \cppinline{gamma=2},
ND in ROOT the yields a result close to $0$ with an absolute error of $10^{-13}$ despite the fact that the function is smooth and well-conditioned at this point. The approximation error becomes larger when the derivative is evaluated further from the critical point. In contrast, the automatic differentiation (in both modes) yields the \emph{exact} result of $0$.

\subsection{Performance}

Section~\ref{background} showed that reverse mode AD computes gradients in a single pass with a runtime complexity of at most $4 \cdot n$, which depends only on the complexity $n$ and not the dimensionality $dim$ of the original function. On the other hand, numerical differentiation requires a separate evaluation of the original function for every dimension to compute the entire gradient, making the overall the run-time complexity of gradient evaluation via central finite difference method $2 \cdot dim \cdot n$. Hence, in theory, reverse mode achieves an asymptotic speedup of $O(dim)$ over the numerical differentiation and can be up to $dim / 2$ times faster.

We experimentally verify this by comparing the performance of gradient evaluation produced by reverse mode AD against our an implementation of numerical differentiation via the central finite difference method. We use the two functions in ~Listing~\ref{lst:sum_mvn}: \cppinline{sum}, which computes the sum of all values in a vector; and \cppinline{mvn}, which implements the PDF of a  multivariate normal distribution. Both functions have a parameter \cppinline{dim} which defines the dimension, and gradients are taken with respect to \cppinline{dim}-dimensional vector~\cppinline{p}. While closed-form expressions of these gradients are well-known, these functions make a good basis of a benchmark as they perform typical operations that are commonly found inside more complicated functions (e.g. \cppinline{+}, \cppinline{*}, \cppinline{pow}, \cppinline{exp} inside loop).

\vspace*{1mm}
\begingroup
\noindent
\begin{minipage}[h]{.45\textwidth}
\begin{cppcode*}{linenos=false}
  double sum(double* p, int dim) {
  double r = 0.0;
  for (int i = 0; i < dim; i++)
    r += p[i];
  return r;
}
\end{cppcode*}
\end{minipage}
\begin{minipage}[h]{.45\textwidth}
\begin{cppcode*}{linenos=false}
  double mvn(double* x, double* p /*means*/,
           double sigma, int dim) {
  double t = 0;
  for (int i = 0; i < dim; i++)
    t += (x[i] - p[i])*(x[i] - p[i]);
  t = -t / (2*sigma*sigma);
  return std::pow(2*M_PI, -n/2.0) *
         std::pow(sigma, -0.5) * std::exp(t);
}
\end{cppcode*}
\end{minipage}
\captionof{listing}{Implementations of \cppinline{sum} and \cppinline{mvn} functions}\label{lst:sum_mvn}
\endgroup

\noindent
Gradients of \cppinline{sum} produced by numerical differentiation and Clad are shown in Listing~\ref{lst:grad_sum}.

\begin{figure}[H]
\centering
\begin{subfigure}[t]{.49\linewidth}
\shadowimage[width=.9\textwidth]{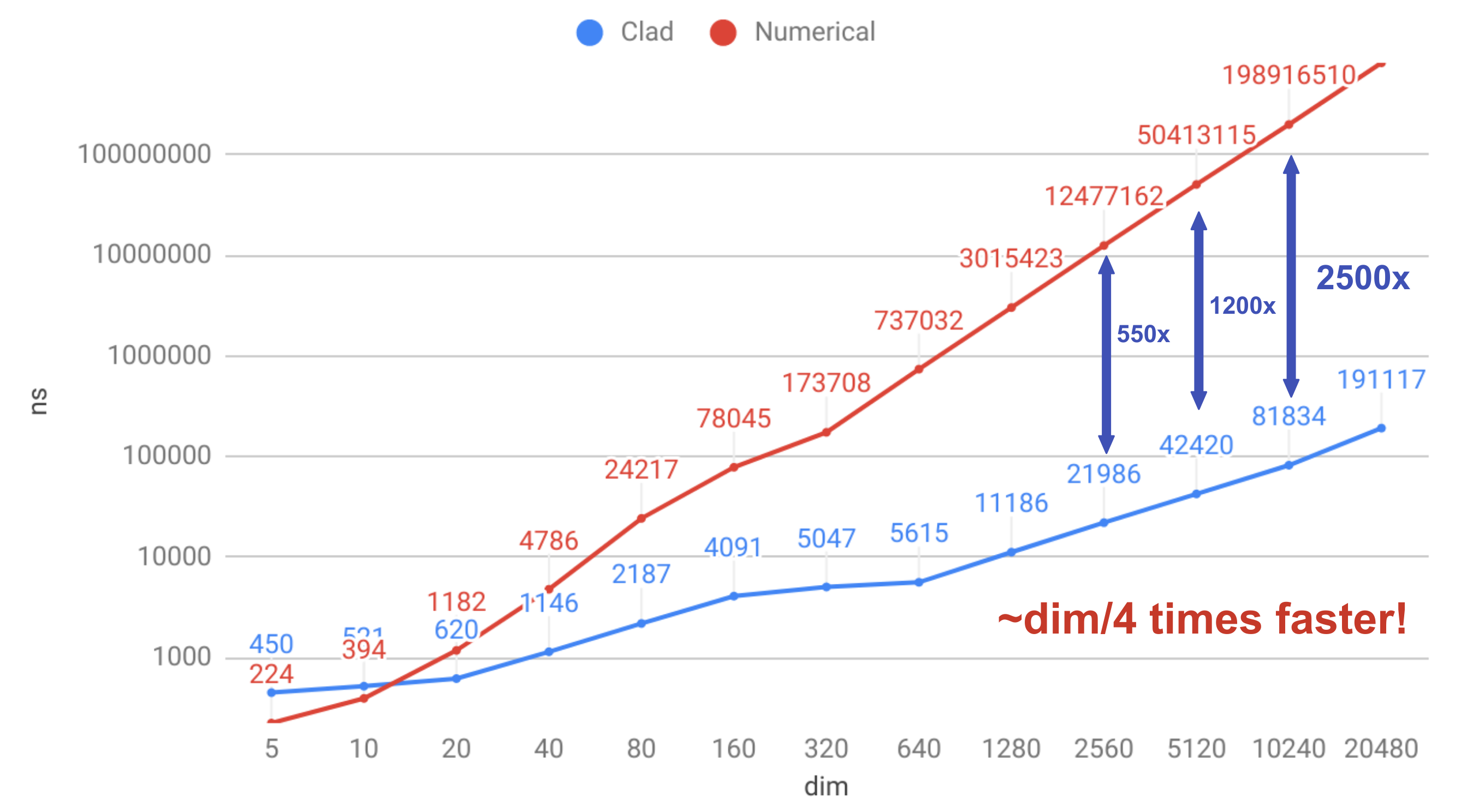}
\caption{Performance of gradients of \cppinline{sum}} \label{fig:perf:c}
\end{subfigure}
\begin{subfigure}[t]{.49\linewidth}
\shadowimage[width=.95\textwidth]{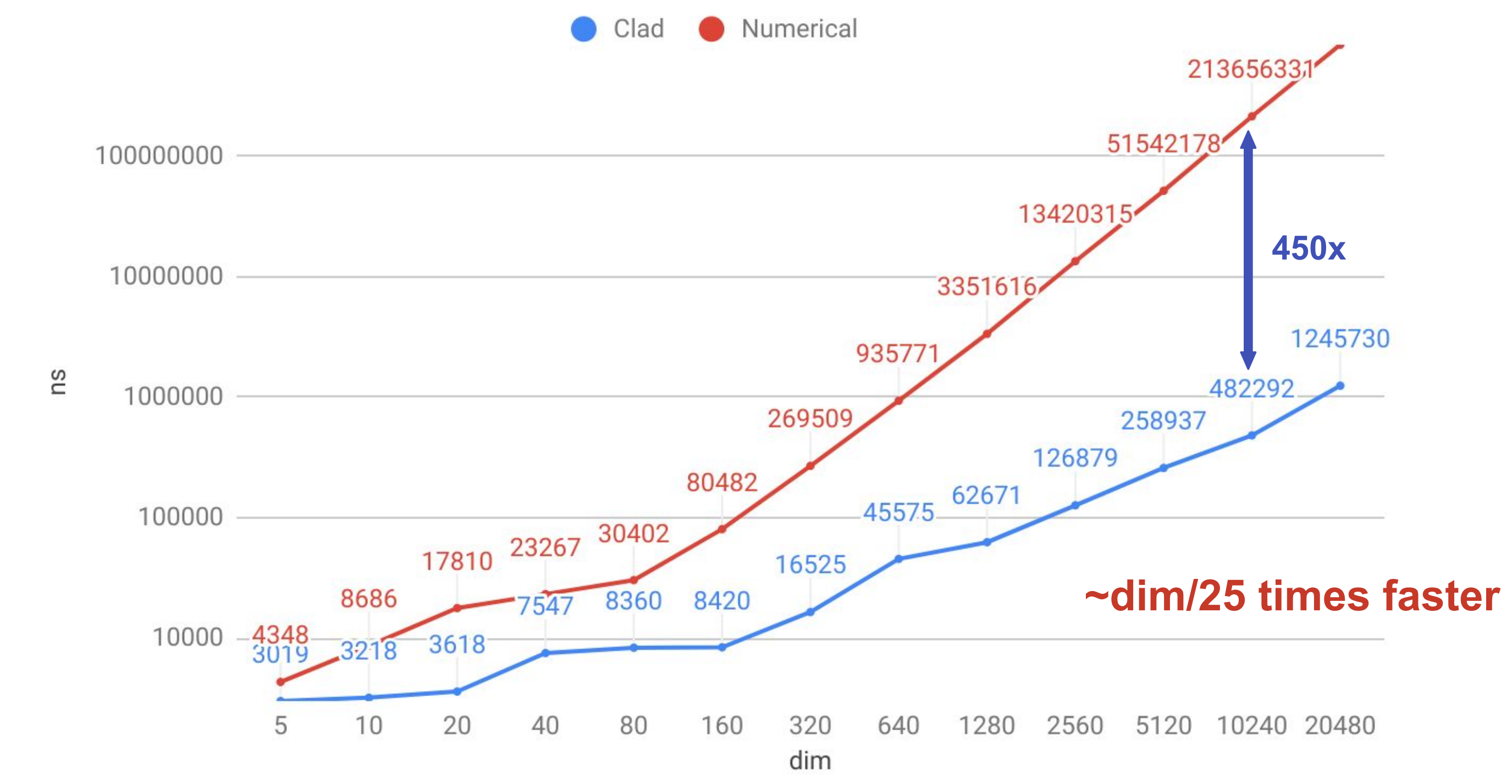}
\caption{Performance of gradients of \cppinline{mvn}} \label{fig:perf:d}
\end{subfigure}
\vspace*{-0.5cm}
\caption{Comparison of reverse mode AD and ND with increasing dimension}\label{fig:performance1}
\end{figure}

\vspace*{1mm}
\begingroup
\noindent
\begin{minipage}[h]{.45\textwidth}
\begin{cppcode*}{linenos=false}
  double* sum_num_grad(double* p, int dim,
                     double eps = 1e-8) {
  double result = new double[dim]{};
  for (int i = 0; i < dim; i++) {
    double pi = p[i];
    p[i] = pi + eps;
    double v1 = sum(p, dim);
    p[i] = pi - eps;
    double v2 = sum(p, dim);
    result[i] = (v1 - v2)/(2 * eps);
    p[i] = pi;
  }
  return result;
}
\end{cppcode*}
\end{minipage}
\noindent
\begin{minipage}[h]{.45\textwidth}
\begin{cppcode*}{linenos=false}
  void sum_ad_grad(double *p, int dim,
              double *_result) {
    double _d_r = 0;
    unsigned long _t0;
    int _d_i = 0;
    clad::tape<int> _t1 = {};
    double r = 0.;
    _t0 = 0;
    for (int i = 0; i < dim; i++) {
        _t0++;
        r += p[clad::push(_t1, i)];
    }
    double sum_return = r;
    _d_r += 1;
    for (; _t0; _t0--) {
        double _r_d0 = _d_r;
        _d_r += _r_d0;
        _result[clad::pop(_t1)] += _r_d0;
        _d_r -= _r_d0;
    }
}
\end{cppcode*}
\end{minipage}
\captionof{listing}{Gradient of \cppinline{sum}: \textbf{(left)} using finite differences, \textbf{(right)} generated by Clad}~\label{lst:grad_sum}
\endgroup
\noindent
We perform the evaluation for values of \cppinline{dim} between $5$ and $20480$. Figure~\ref{fig:performance1} shows the comparison for \textbf{(a)} \cppinline{sum}; \textbf{(b)} \cppinline{mvn} and confirms the expected theoretical speedup of $O(dim)$, with AD-generated gradient being $~dim/4$ times faster for \cppinline{sum} and $~dim/25$ times faster for \cppinline{mvn} (slowdown is due to more expensive operations like \cppinline{pow}, \cppinline{exp}).


\vspace*{-0.4cm}
\vspace*{-1cm}

\subsection{Performance in \emph{TFormula}} \label{perftformula} 
\begin{figure}[h!]
\centering
\begin{subfigure}[t]{.49\linewidth}
  \shadowimage[width=.9\textwidth]{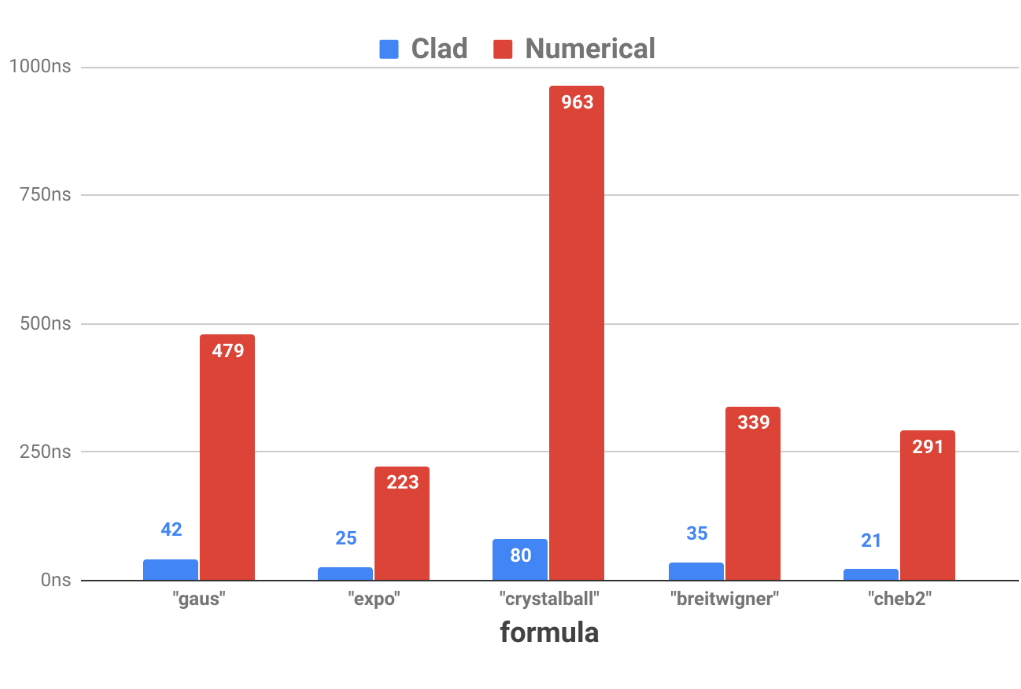}
  \caption{Comparison of the performance of  \cppinline{TFormula} gradients between AD and ND}
  \label{fig:perf:a}
\end{subfigure}
\begin{subfigure}[t]{.49\linewidth}
  \shadowimage[width=.9\textwidth]{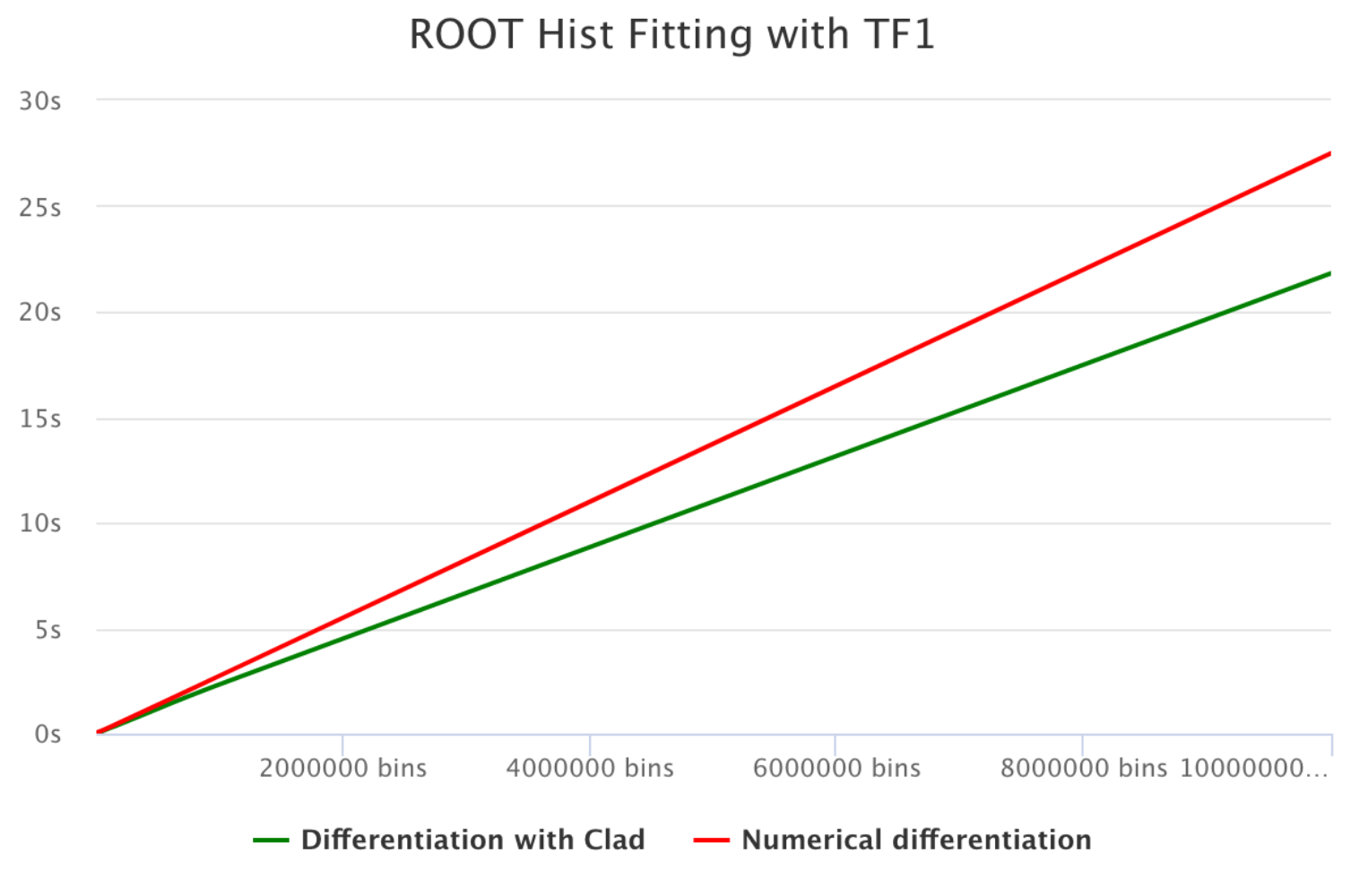}
  \caption{ROOT Histogram fitting using numerical differentiation versus Clad differentiation}
  \label{fig:perf:b}
\end{subfigure}
\caption{Performance benchmarks in ROOT}
\end{figure}

Figure~\ref{fig:perf:a} shows the performance comparisons of reverse-mode AD and ND for the task of evaluating gradients of \cppinline{TFormula}'s builtin primitive probability density functions. The functions are \cppinline{gaus}~($dim=3$), \cppinline{expo}~($dim=2$), \cppinline{crystalball}~($dim=5$), \cppinline{breitwigner}~($dim=5$) and \cppinline{cheb2}~($dim=4$). Despite the low dimensionality ($dim \leq 5$), AD gives significant (approx. 10x) speedups. The speedups are even larger than expected factor of $dim/2$ that follows from theoretical results, apparently due to additional overhead of the implementation of numerical differentiation in ROOT, which tries to find the optimal step size for its finite difference method to improve accuracy.

In Figure~\ref{fig:perf:b}, we perform fitting of a Gaussian distribution to a histogram of random samples via gradient-based optimization. In ROOT, this functionality is implemented in \cppinline{TFormula}-based \cppinline{TF1} class. We can therefore use AD due to the integration of Clad into \cppinline{TFormula}.
Figure~\ref{fig:perf:b} compares the performance of the AD-based \cppinline{TF1} fitting with the numerical fitting in the \cppinline{Hist} package. As in previous experiments, we show that AD scales better with problem dimensionality (number of histogram bins) on this task.  The integration of Clad into \cppinline{TFormula} makes it straightforward to use AD for fitting in ROOT.

\section{Conclusion} \label{conclusion}
We discussed our implementation of automatic differentiation in ROOT based on Clad. We demonstrated that Clad is integrated into ROOT and can be easily used in various contexts inside ROOT (e.g. histogram fitting).
Furthermore, we showed that automatic differentiation in ROOT achieves significant improvements in accuracy and performance over numerical differentiation. The performance and accuracy are promising and encourage further work in the development of Clad and its integration in ROOT.



Possible further improvements for Clad include optimizations to code transformation and design of a consistent interface for derivatives and gradients computation. This functionality can be further extended, including the computation of Jacobians and higher-order derivatives. In order to achieve optimal performance, the evaluation of individual derivatives could be executed in parallel. Besides, the Clad API should enable a flexible execution method based on the needs of its user.

\section{Acknowledgments}

This work has been supported by U.S. NSF grants PHY-1450377 and 1450323.

\bibliography{main}

\begin{thebibliography}{17}

\bibitem{baydin2017automatic}
A.G. Baydin, B.A. Pearlmutter, A.A. Radul, J.M. Siskind, \emph{Automatic
  differentiation in machine learning: a survey}, in \emph{The Journal of
  Machine Learning Research} (JMLR.org, 2017), Vol.~18, pp. 5595--5637

\bibitem{gebremedhin2020introduction}
A.H. Gebremedhin, A.~Walther, \emph{An introduction to algorithmic
  differentiation}, in \emph{Wiley Interdisciplinary Reviews: Data Mining and
  Knowledge Discovery} (Wiley Online Library, 2020), Vol.~10, p. e1334

\bibitem{griewank2008evaluating}
A.~Griewank, A.~Walther, \emph{Evaluating derivatives: principles and
  techniques of algorithmic differentiation}, Vol. 105 ({SIAM}, 2008)

\bibitem{Brun1997ROOT}
R.~Brun, F.~Rademakers, \emph{{ROOT} {\textemdash} An object oriented data
  analysis framework}, in \emph{Nuclear Instruments and Methods in Physics
  Research Section A: Accelerators, Spectrometers, Detectors and Associated
  Equipment} (Elsevier {BV}, 1997), Vol. 389, pp. 81--86

\bibitem{vassilev2015clad}
V.~Vassilev, M.~Vassilev, A.~Penev, L.~Moneta, V.~Ilieva, \emph{Clad --
  automatic differentiation using Clang and LLVM}, in \emph{Journal of Physics:
  Conference Series} (IOP Publishing, 2015), Vol. 608, p. 012055

\bibitem{cladgithub}
V.~Vassilev, \emph{Clad -- automatic differentiation for {C/C}++} (2014),
  \urlstyle{tt}\url{https://github.com/vgvassilev/clad/}

\bibitem{bischof1996adifor}
C.~Bischof, P.~Khademi, A.~Mauer, A.~Carle, \emph{ADIFOR 2.0: Automatic
  differentiation of Fortran 77 programs}, in \emph{IEEE Computational Science
  and Engineering} (IEEE, 1996), Vol.~3, pp. 18--32

\bibitem{bischof1997adic}
C.H. Bischof, L.~Roh, A.J. Mauer-Oats, \emph{ADIC: an extensible automatic
  differentiation tool for ANSI-C}, in \emph{Software: Practice and Experience}
  (Wiley Online Library, 1997), Vol.~27, pp. 1427--1456

\bibitem{paszke2017automatic}
A.~Paszke, S.~Gross, S.~Chintala, G.~Chanan, E.~Yang, Z.~DeVito, Z.~Lin,
  A.~Desmaison, L.~Antiga, A.~Lerer, \emph{Automatic differentiation in
  PyTorch} (2017)

\bibitem{jax2018github}
J.~Bradbury, R.~Frostig, P.~Hawkins, M.J. Johnson, C.~Leary, D.~Maclaurin,
  S.~Wanderman-Milne, \emph{{JAX}: composable transformations of
  {P}ython+{N}um{P}y programs} (2018),
  \urlstyle{tt}\url{http://github.com/google/jax}

\bibitem{innes2019differentiable}
M.~Innes, A.~Edelman, K.~Fischer, C.~Rackauckas, E.~Saba, V.B. Shah,
  W.~Tebbutt, \emph{A differentiable programming system to bridge machine
  learning and scientific computing} (2019)

\bibitem{swiftfortensorflow}
\emph{Swift for {TensorFlow}},
  \urlstyle{tt}\url{https://www.tensorflow.org/swift}

\bibitem{autodiffcpp}
M.~Pulver, \emph{Autodiff -- automatic differentiation c++ library} (2019),
  \urlstyle{tt}\url{https://github.com/pulver/autodiff}

\bibitem{hogan2014fast}
R.J. Hogan, \emph{Fast reverse-mode automatic differentiation using expression
  templates in C++}, in \emph{ACM Transactions on Mathematical Software (TOMS)}
  (ACM New York, NY, USA, 2014), Vol.~40, pp. 1--16

\bibitem{walther2009getting}
A.~Walther, A.~Griewank, \emph{Getting Started with ADOL-C.}, in
  \emph{Combinatorial scientific computing} (2009), 09061, pp. 181--202

\bibitem{LLVM:CGO04}
C.~Lattner, V.~Adve, \emph{{LLVM}: A Compilation Framework for Lifelong Program
  Analysis and Transformation} (San Jose, CA, USA, 2004), pp. 75--88

\bibitem{Vasilev2012Cling}
V.~Vasilev, P.~Canal, A.~Naumann, P.~Russo, \emph{{Cling -- The New Interactive
  Interpreter for ROOT 6}}, in \emph{Journal of Physics: Conference Series}
  (IOP Publishing, 2012), Vol. 396, p. 052071

\end{thebibliography}

\end{document}